\documentstyle[prl,aps,epsfig]{revtex}         
\begin{document}
\catcode`\ä = \active \catcode`\ö = \active \catcode`\ü = \active
\catcode`\Ä = \active \catcode`\Ö = \active \catcode`\Ü = \active
\catcode`\ß = \active \catcode`\é = \active \catcode`\è = \active
\catcode`\ë = \active \catcode`\ô = \active \catcode`\ê = \active
\catcode`\ø = \active \catcode`\ò = \active \catcode`\í = \active
\defä{\"a} \defö{\"o} \defü{\"u} \defÄ{\"A} \defÖ{\"O} \defÜ{\"U} \defß{\ss} \defé{\'{e}}
\defè{\`{e}} \defë{\"{e}} \defô{\^{o}} \defê{\^{e}} \defø{\o} \defò{\`{o}} \defí{\'{i}}
\draft               
\twocolumn[\hsize\textwidth\columnwidth\hsize\csname
@twocolumnfalse\endcsname
\title{Vortex Nucleation in a Stirred Bose-Einstein Condensate}
\author{C. Raman, J. R. Abo-Shaeer, J. M. Vogels, K. Xu, and W. Ketterle}
\address{Department of Physics, MIT-Harvard Center for Ultracold
Atoms, and Research Laboratory
of Electronics, \\
MIT, Cambridge, MA 02139}
\date{\today}
\maketitle
\begin{abstract}
We studied the nucleation of vortices in a Bose-Einstein
condensate stirred by a laser beam.  We observed the vortex cores
using time-of-flight absorption imaging. By varying the size of
the stirrer, we observed either discrete resonances or a broad
response as a function of the frequency of the stirrer's motion.
Stirring beams small compared to the condensate size generated
vortices below the critical rotation frequency for the nucleation
of surface modes, suggesting a local mechanism of generation.  In
addition, we observed the centrifugal distortion of the condensate
due to the rotating vortex lattice and found evidence for bent
vortices.
\end{abstract}
 \vskip1pc ]

Dissipation and turbulence in superfluid flow often involves the
creation and subsequent motion of quantized vortices
\cite{donn91}.  Since vortices are topological defects they may
only be created in pairs, or can enter a system individually from
its boundary. The nucleation process has been a subject of much
theoretical interest \cite{fett01}. Experiments with Bose-Einstein
condensates (BEC) confined in atom traps are well suited to test
theories of nucleation because the boundary of the condensate is
well controlled, and vortices can be directly imaged.

Vortex nucleation usually involves dynamical instabilities and
superfluid turbulence \cite{donn91,fett01}.  Exceptions are the
direct coupling between ground and vortex states for small
condensates \cite{cara99} and the engineering of states of
quantized circulation by manipulating the phase of the
wavefunction \cite{matt99vort,dobr99,marz98}. Turbulent flow can
be created by perturbing the system with a time-dependent
boundary, for example by a small laser beam
\cite{rama99,inou01vort}, or by a rotating trap anisotropy
\cite{madi00,abos01latt,halj01damop}.  The resulting vortices have
been directly imaged \cite{inou01vort,madi00,abos01latt}.

In a rotating frame, vortices are energetically favored above a
critical rotational frequency $\Omega_c$ \cite{fett01,nozi90}.
Current theories suggest that the nucleation of vortices requires
the creation of surface waves which ``collapse'' into a vortex
state \cite{dalf97coll,isos99,alkh99,dalf01crit,fede01}. This
corresponds to the disappearance of the energy barrier for a
vortex to enter the cloud \cite{fett01,isos99,fede99A}.  For
surface excitations with angular momentum $l\hbar$ and frequency
$\omega_l$, the minimum rotational frequency is \vspace{-0.3 cm}
\begin{equation}
\Omega_s= {\rm min}_{l}(\omega_l/l). \label{eq:surface}
\end{equation}
This corresponds to a Landau critical velocity for surface waves
$v_c = \Omega_s R_{\rm TF}$, where $R_{\rm TF}$ is the
Thomas-Fermi radius \cite{alkh99}.

Our main result is that vortices can be created {\em below}
$\Omega_s$ by using a small, localized stirring beam.  This
indicates that the current surface mode analysis of vortex
nucleation is incomplete. Recent experiments \cite{madi00,madi01}
observed vortices only at much \emph{higher} frequencies, giving
rise to a variety of theoretical models \cite{fede01,garc00vort}.
One explanation is that those experiments excited only the $l=2$
mode, which requires a higher drive frequency than Eqn.
(\ref{eq:surface}) \cite{madi01,sinh01prep,chev01high_l}.  We
verified this prediction by stirring the condensate with
anisotropies of different symmetries ($l=2, 3, 4$) and observing
distinct resonance frequencies for vortex formation.  If the
condensate is stirred with a small beam, one would expect to
couple to many modes $l$.  Surprisingly, the small stirrer did not
excite resonances, but could nonetheless generate vortices as
effectively as a resonant drive.

Our method of vortex generation has been outlined in previous work
\cite{abos01latt}.  We start with nearly pure BECs ($> 90 \%$
condensate fraction) of up to 5$\times10^7$ sodium atoms in a
cylindrically shaped Ioffe-Pritchard magnetic trap with radial and
axial frequencies of $\omega_r = 2\pi \times$ 83 Hz and $\omega_z
= 2\pi \times$ 20 Hz, respectively.  A radio frequency ``shield''
limited the magnetic trap depth to 50 kHz (2.3 $\mu$K). The
condensate chemical potential and peak density were 300 nK and $4
\times 10^{14}$ cm$^{-3}$, respectively, corresponding to a
healing length $\xi \simeq 0.2$ $\mu{\rm m}$.

\begin{figure}[htbf]
\begin{center}
\epsfxsize=70mm {\epsfbox{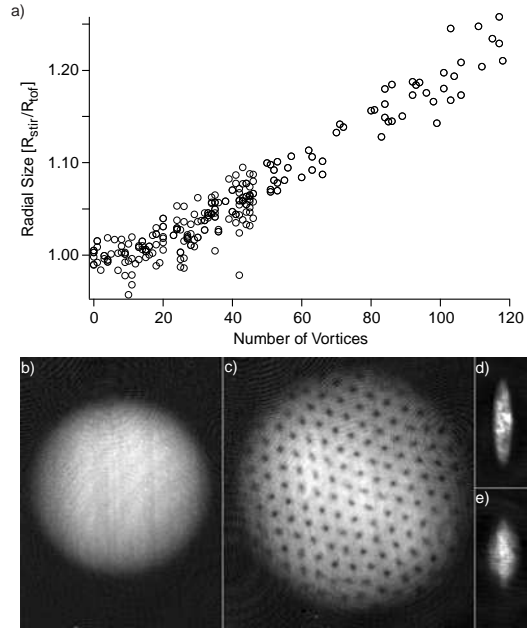}}
\end{center}
\caption{Centrifugal effects on a rotating condensate.  Shown is
the observed radial size of the condensate versus vortex number.
The condensate size was determined from a parabolic fit to the
time-of-flight distribution after 41.5 ms. (b,c) Comparison of a
non-rotating condensate ($R_{\rm tof}$ = 570 $\mu$m) and one with
160 vortices.  Radial in-situ phase contrast imaging of a
condensate at rest (d) and in rotation at a frequency of $~ 75$ Hz
(e), showing the modified aspect ratio. The length of the
condensate in (d) was 200 $\mu$m.}\label{fig:RadiusVsNumber}
\end{figure}

Vortices were generated by rotating the condensate along its long
axis with a scanning blue-detuned laser beam (532 nm)
\cite{onof00}. We explored different beam waists between 5 and 25
$\mu$m.  For the tightest focus, the peak optical dipole potential
was 620 nK.  We used scan radii as large as the Thomas-Fermi
radius $R_{\rm TF}$, which varied from 27-30 $\mu$m. Multiple beam
patterns (formed by rapidly scanning the laser beam from 1.5-10
kHz) were used to create vortices. The laser beam was left on
during evaporation to damp out dipole motion of the condensate.
Immediately after producing a condensate we began the rotation for
times of up to 500 ms, generating a vortex tangle. The laser beams
were then instantly shut off and the cloud equilibrated for 500
ms, during which time the vortices crystallized into an Abrikosov
lattice as shown in Fig. \ref{fig:RadiusVsNumber}c and detailed in
previous work \cite{abos01latt}. For small numbers of vortices the
gas did not fully settle into a regular lattice before imaging.

The vortex cores were observed using resonant absorption imaging
after 41 ms of ballistic expansion, which magnified them by 20
from their size $\xi$ in the trap.  As in our previous work we
imaged a 50-150 $\mu$m slice of atoms in the center of the cloud
using spatially selective optical pumping into the $F=2$ to $F=3$
cycling transition \cite{abos01latt}.
\begin{figure}[htbf]
\begin{center}
\epsfxsize=70mm {\epsfbox{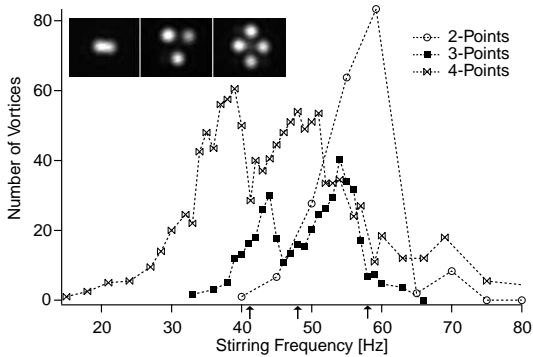}}
\end{center}
\vspace{-1.7cm} \caption{Discrete resonances in vortex nucleation.
The number of vortices created by multi-point patterns is shown.
The condensate radius was R$_{\rm TF}$ = 28~$\mu$m. Each data
point is the average of three measurements.  The arrows below the
graph show the positions of the surface mode resonances $\omega_l
= \omega_r/\sqrt{l}$. The stirring times were 100 ms for the 2-
and 3-point data, and 300 ms for the 4-point data.  Inset shows
2-,3-, and 4-point dipole potentials produced by a 25 $\mu$m waist
laser beam imaged onto the CCD camera.  The separation of the
beams from the center is 25~$\mu$m for the 2-point pattern and
55~$\mu$m for the 3- and 4-point patterns. The laser power per
spot was 0.35 mW, 0.18 mW, and 0.15 mW for the 2-,3- and 4-point
data, respectively.
 } \label{fig:NPointRotations}
\end{figure}
By varying the stirring parameters we explored different
mechanisms for vortex nucleation. A large stirrer, with a beam
waist comparable to the Thomas-Fermi radius showed enhanced vortex
generation at discrete frequencies. Fig.~\ref{fig:NPointRotations}
shows the number of vortices versus the frequency of rotation of
the laser beam using 2-, 3- and 4-point patterns.  The total laser
beam power corresponded to an optical dipole potential between 60
nK and 240 nK. The resonances were close to the frequencies of
excitation of $l=$2, 3 and 4 surface modes
($\omega_l=\omega_{r}/\sqrt{l}$) \cite{dalf97coll}.  A second,
higher resonance appeared in the 3- and 4-point data, presumably
due to additional coupling to the quadrupole ($l=2$) mode.  The
shift in the position of the peaks from the frequencies $\omega_l$
of elementary excitations is probably due to the presence of the
stirrer and the vortices.

Our results clearly show discrete resonances in the nucleation
rate of vortices which depend on the geometry of the rotating
perturbation.  This confirms the role of discrete surface modes in
vortex formation.  A dependence on the symmetry of the stirrer
(1-point versus 2-point) has also been explored in Paris
\cite{chev01high_l}.  For longer stirring times and higher laser
powers the condensate accommodated more vortices at all
frequencies, and the resonances became less pronounced.

\begin{figure}[htbf]
\begin{center}
\epsfxsize=70mm{\epsfbox{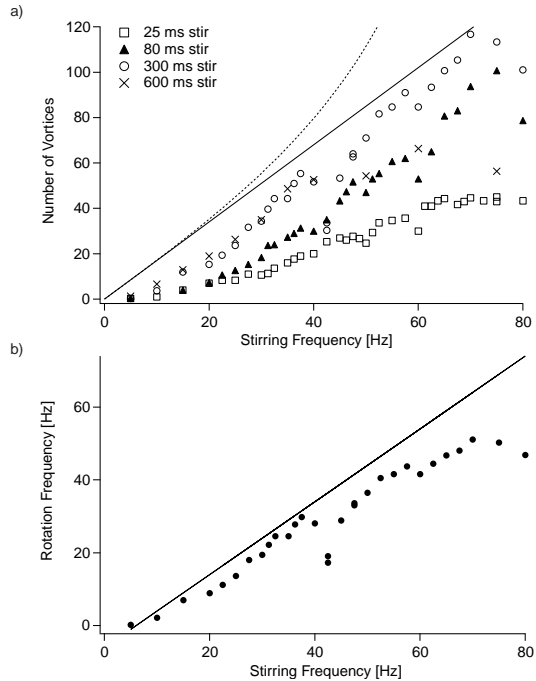}}
\end{center}
\caption{Nonresonant nucleation using a small stirrer.  (a)
Average number of vortices created for different stirring times
using a 2-point pattern positioned at the edge of the condensate.
The beam waist, total power, and separation are 5.3~$\mu$m, 0.16
mW, and 54~$\mu$m, respectively. (b)  Effective lattice rotation
frequency as a function of stir frequency.  The lines in both
graphs indicate the predictions of different models described in
the text.}\label{fig:smallbeam}
\end{figure}

A stirrer much smaller than the condensate size could generate
vortices very rapidly -- more than 100 vortices were created in
100 ms of rotation.  Fig.~\ref{fig:smallbeam}a shows the number of
vortices produced using a 2-point pattern with a scan radius close
to R$_{\rm TF}$ for various stirring times.  Above 300 ms the
angular momentum of the cloud appeared to saturate and even
decreased, accompanied by visible heating of the cloud.
The maximum number of vortices was roughly proportional to the
stirring frequency.  For a lattice rotating at frequency $\Omega$,
the quantized vortex lines are distributed with a uniform area
density $n_v = \frac{2 \Omega}{\kappa}$, where $\kappa = h/M$ is
the quantum of circulation and $M$ the atomic mass \cite{nozi90}.
Therefore, the number of vortices at a given rotation frequency
should be
\begin{equation}
N_v = 2 \pi R^2 \Omega/\kappa \label{eq:number}
\end{equation}
in a condensate of radius $R$.  The straight line in Fig.
\ref{fig:smallbeam}a assumes $R=R_{\rm TF}$ and that the lattice
has equilibrated with the drive.  In contrast to the large stirrer
no resonances were visible even when the number of vortices had
not yet saturated.  This suggests a different mechanism for vortex
nucleation.  Further evidence for this was obtained from the
frequency and spatial dependences.

For our experimental conditions, Eqn.~\ref{eq:surface}) yields
$\Omega_s \simeq 0.25 \omega_r = 21$ Hz \cite{fede01priv}.  With
the small stirrer, we observed vortices at frequencies as low as 7
Hz. Below this frequency the velocity of the stirrer was not much
larger than the residual dipole oscillation of the condensate. The
rotational frequency below which a rectilinear vortex in the
condensate center is energetically favored is 7 Hz \cite{fett01}.

\begin{figure}[htbf]
\begin{center}
\epsfxsize=70mm {\epsfbox{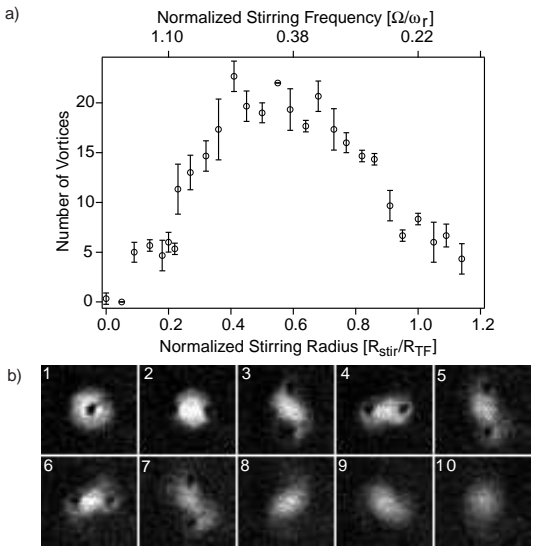}}
\end{center}
 \caption{Spatial dependence of vortex nucleation.  (a) Shown is
the number of vortices as a function of the stirring radius for a
constant stirring velocity (0.2 $c$). At the Thomas-Fermi radius,
the angular frequency of the stirrer was 17.5 Hz.  The error bars
reflect the scatter of 3 measurements at the same radius. (b)
Axial phase contrast images showing the pure condensate with the
aligned stirrer (1), the condensate during (2-7) and after (8-10)
stirring.} \label{fig:Position}
\end{figure}

In Fig.~\ref{fig:Position} we varied the radius of the 2-point
scan. The frequency of rotation was chosen to keep the linear
velocity of the laser beam constant. Vortices could be generated
over a broad range of radii.  The maximum number was obtained at
intermediate radii rather than the Thomas-Fermi surface, providing
further evidence that surface excitations of the unperturbed
condensate are not the dominant nucleation mechanism.   The
observed radial dependence makes it unlikely that the thermal
cloud plays a crucial role in the vortex nucleation, since at very
low temperatures its maximum density occurs at the surface of the
condensate.  Indeed, we observed that finite temperatures
increased the damping and led to fewer vortices.
The nucleation mechanism for small stirrers may be related to our
earlier experiments on the onset of dissipation in stirred
condensates, where we observed a drag force at velocities above
$\sim 0.1 c$, where $c$ is the speed of sound at the condensate
center \cite{rama99,onof00sup}.  The friction with the moving
stirrer causes an asymmetry in the density profile in front of and
behind the laser beam.  This has been directly imaged for linear
motion \cite{onof00sup}.  Similar flow field effects can be
observed in Fig. \ref{fig:Position}b, where they are clearly
linked to the formation of vortices.  Vortex pairs are predicted
to arise from linear stirring \cite{fris92,jack99}. When the laser
beam moves in a circle, co-rotating vortices will be favored,
whereas counter-rotating vortices will be expelled from the
system.

For an object smaller than the healing length $\xi$, the critical
velocity for vortex formation occurs at the Landau value given by
Eqn. (\ref{eq:surface}). For objects larger than $\xi$ such as our
laser beam, hydrodynamic flow around the stirrer can reduce the
critical velocity relative to Eqn. (\ref{eq:surface})
\cite{fris92}, which may explain our observation of vortices below
21 Hz.

In our earlier work we observed that the size of a condensate with
vortices exceeded the size of the non-rotating condensate
\cite{abos01latt}.  Here we study these centrifugal distortions
quantitatively (also note \cite{halj01damop}).
Fig.~\ref{fig:RadiusVsNumber} shows the enhancement of the cloud
size $R$ in time-of-flight by up to 25\% due to additional
rotational kinetic energy.  The rotating condensate experiences a
centrifugal potential $-(1/2)M (\Omega r)^2$, which leads to an
effective radial trapping frequency of $\sqrt{\omega_r^2 -
\Omega^2}$.  For a constant number of atoms $N$, this increases
the Thomas Fermi radius
\begin{equation}
R_{\rm TF} = \frac{R_0}{(1-(\frac{\Omega}{\omega_r})^2)^{3/10}}
\label{eq:size}
\end{equation}
and reduces the mean-field interaction energy $E_{int}$. The total
release energy of the gas is then $E = E_{int} + \frac{1}{2}
I_{\rm eff} \Omega^2 + E_{v0} N_v$. The 2nd term accounts for the
rigid rotation of the lattice while the 3rd term is a quantum
correction due to the kinetic energy of the cores, and is
negligible for large $N_v$.  The effective moment of inertia of
the condensate is $I_{\rm eff} = 2/7 M R_{\rm TF} ^2 N$. The
energy per unit length of a single vortex is $E_{v0} = \pi n
\hbar^2/M \ln{[0.27 d/\xi]}$, where $d = n_v^{-1/2}$ is the mean
distance between vortices \cite{stau68}. When we average $E_{v0}$
over the Thomas-Fermi distribution we predict a $30\%$ increase in
$E$ for 120 vortices, whereas the observed increase in $R^2$ is
about $50\%$. This discrepancy is probably due to our selection of
the central slice of the cigar, where the rotational energy is the
highest, and due to the unobserved axial expansion, which may
depend on the angular momentum of the cloud.

If we account for centrifugal effects and combine Eqns.
(\ref{eq:number}) and (\ref{eq:size}), we expect a divergence of
the number of vortices near the trap frequency (dashed line in
Fig. \ref{fig:smallbeam}a).  The deviation of the data from this
line suggests that the condensate did not fully equilibrate with
the rotating drive.  Taking into account the critical velocity for
vortex nucleation, $v_c \simeq 0.1 c$ \cite{onof00sup}, we expect
the maximum rotation frequency of the lattice to be $\Omega_S -
v_c/R_{\rm TF}$ where $\Omega_S$ is the frequency of the stirrer
moving at radius $R_{\rm TF}$. Using the measured number of
vortices, we can invert Eqns. (\ref{eq:number}) and
(\ref{eq:size}) to derive the lattice rotation frequency, which is
shown in Fig. \ref{fig:smallbeam}b, along with the expected value
$\Omega_S - 2 \pi \times 6$ Hz assuming a constant $N$. The
discrepancy can be partly attributed to loss of atom number due to
heating by the stirrer, which was up to 30\%.  We can also derive
from these equations the flow velocity at the edge of the
condensate.  For a lattice with 144 vortices, this velocity
exceeded the speed of sound at the condensate center by 40\%, in
contrast to a recent suggestion that supersonic rotation speeds
are unattainable\cite{fedi00}.

\begin{figure}[htbf]
\begin{center}
\epsfxsize=70mm {\epsfbox{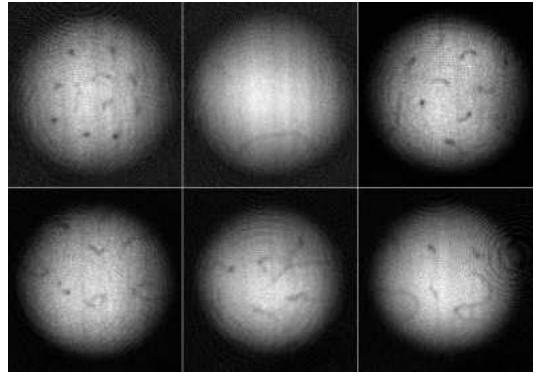}}
\end{center}
\caption{
 Three-dimensional structure of vortices.  Shown are
several examples of time-of-flight pictures of condensates at low
rotational frequencies, where "smeared-out" vortex cores and
elongated features were observed.  The radius of the condensate
was 510 $\mu$m.} \label{fig:BentVortices}
\end{figure}

At low rotational velocities, vortices should not be rectilinear
as assumed in many theoretical calculations but bent
\cite{garc01bend,afta01}. Such bent vortices should have lower
visibility in our images due to the line of sight integration
across the optically pumped condensate slice.
Fig.~\ref{fig:BentVortices} shows several examples.  Some appear
as vortex lattices with tilted vortex cores.  Other images show
structures reminiscent of half rings and coiled vortices. However,
since these are time-of-flight images it is not obvious how some
of the observed features are related to spatial
structures in the trapped condensate. 

In conclusion, we have studied the nucleation of vortices in
condensates stirred by rotating laser beams and identified two
different nucleation mechanisms, surface modes for large stirrers,
as well as local turbulence for small stirrers.  So far, we have
mainly concentrated on pure condensates, but it will be intriguing
to study the vortex phase diagram at finite temperatures
\cite{stri99vort,mizu01} and the role of the thermal cloud in the
decay of vortices\cite{fedi00}.

We thank J.R. Anglin, D. Feder, A. E. Leanhardt and A.P. Chikkatur
for useful discussions. This research is supported by NSF, ONR,
ARO, NASA, and the David and Lucile Packard Foundation.
\vspace{-.3cm}
\vspace{-1.2 cm}

\end{document}